# Camouflage Design of Analysis Based on HSV Color Statistics and K-means Clustering

**Xinyu Wei, Mengjia Zhou, Bernie Liu**

**Abstract:**

Since ancient times, it has been essential to adopting camouflage on the battlefield, whether it is in the forefront, in depth or the rear. Traditional evaluation method is made up of people’s opinion. By watching target or looking at the pictures, and determine the effect of camouflage, so it can be more influenced by man's subjective factors. And now, in order to objectively reflect the camouflage effect, we set up model through using image’s similarity to evaluate camouflage effect. Image similarity comparison is divided into two main image feature comparison: image color features and texture features of images.

We now using computer design camouflage, camouflage pattern design is divided into two aspects of design color and design plaques. For the design of the color, we based on HSV color model, and as for the design of plague, the key steps is the background color edge extraction, we adopt algorithm based on k-means clustering analysis of the method of background color edge extraction.

# 1 Introduction

## 1.1 Problem description

In the traditional design of camouflage, camouflage color and spots is mainly composed of people with experience in design, not only waste money and time, and the efficiency is low, cannot achieve good camouflage effect.Therefore artificial design does not improve the camouflage effect of camouflage, resulting in a decline in the operational capability of the army.In modern military, good camo is particularly important, our goal is to design a can adapt to most environments (such as forest land, sand, desert and snow) camouflage, and establish an effective evaluation model, and developed a can quickly and accurately on the computer design of camouflage methods, improve the camouflage effect of camouflage.

### 1.2 Assumption

(1) Assuming four major combat fields are: forest land, sand, snow and desert

(2) Assuming the results of camouflage are mainly affected by color and patch, ignoring other factors, such as light.

(3) Assuming when the enemy detective the environment, it is not affected by noise.

## 2. Model analysis

### 2.1 Process of solving problems

(1 )Establish the model evaluation from the aspects of color and texture;

(2) Extract the background color, the color of camouflage design;

(3) Extract the rough outline of the background, design the spot shape of the camouflage

(4) Evaluate the design of camouflage

(5) Design camouflage advertisements.

### 2.2 The major factors for the camouflage effective

(1) The protection color of the camouflage

(2) The shape of the camouflage
(3) Camouflage spot size

# 2 The camouflage color evaluation model based on color histogram

When we camouflaging on military target, the effect of camouflage determines the probability whether the place could be found. So it's important to establish a model that can evaluate the results of different camouflages. Traditional evaluation method is made up of people's opinion. By watching target or looking at the pictures, and determine the effect of camouflage, so it can be more influenced by man's subjective factors. And now, in order to objectively reflect the camouflage effect, we set up model through using image's similarity to evaluate camouflage effect.

Image similarity comparison is divided into two main image feature comparison: image color features and texture features of images. Determines whether the image similarity is through the similar characteristics, which is using the image similarity comparison to show the comparison between the image features. For image color feature similarity detection, we use the image color based on color histogram similarity detection method.

## 2.3 The detection of similarity of the image color feature

### 2.3.1 The definition of a histogram

Assume SX(i) isthenumbers of acharacteristic value in image P, N Is the total number of pixels in the image P, now we do the normalized processing with SX(i) , which is:

$$h(X_i) = \frac{S(X_i)}{N} = \frac{S(X_i)}{\sum_j S(X_i)}$$

So the histogram of the image is

$$H(P) = \left(h(X_1), h(X_2), \ldots, h(X_n)\right)$$

### 2.3.2 Based on the principle of the detection of histogram color similarity

Bhattacharyya distance: For discrete probability distributions p and q over the same domain X, it is defined as:

$$D_B(p,q) = -\ln\left(BC(p,q)\right)$$

where:

$$BC(p,q) = \sum_{x \in X} \sqrt{p(x)q(x)}$$

is the Bhattacharyya coefficient.

For continuous probability distributions, the Bhattacharyya coefficient is defined as:

$$BC(p,q) = \int \sqrt{p(x)q(x)}\, dx$$

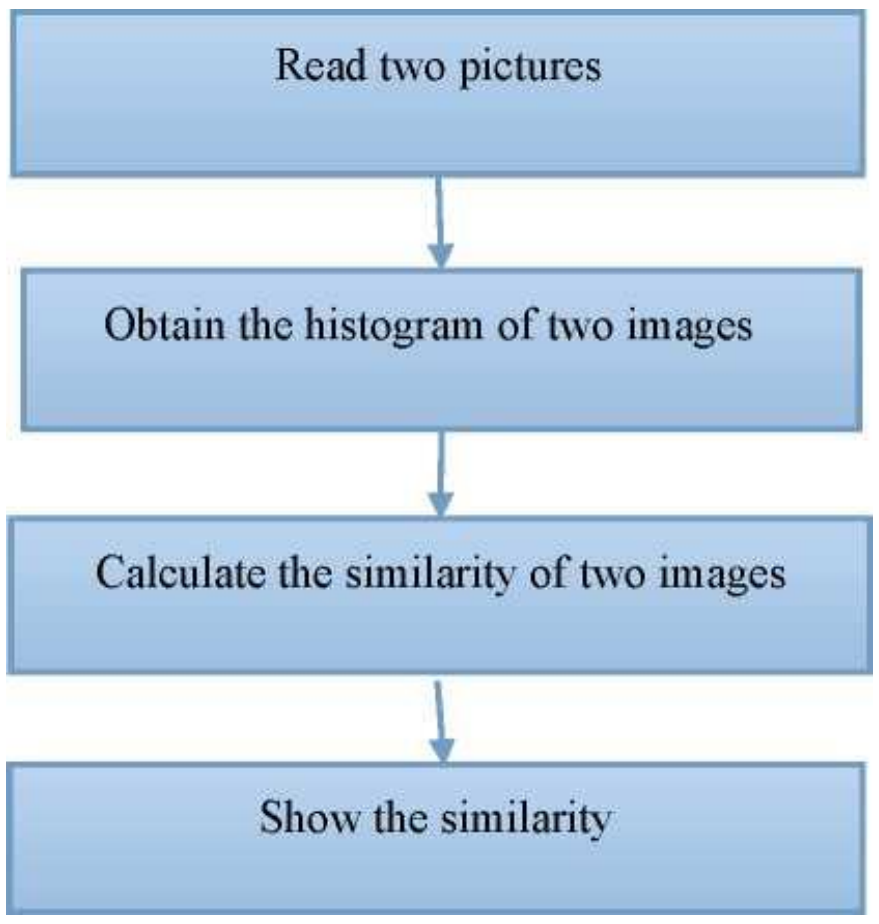

**Figure 1 The process of Histogram of color detection of the model**

### 3.1.4 The effect of the model of histogram testing existing camouflage

Woodland camouflage camouflage effect test

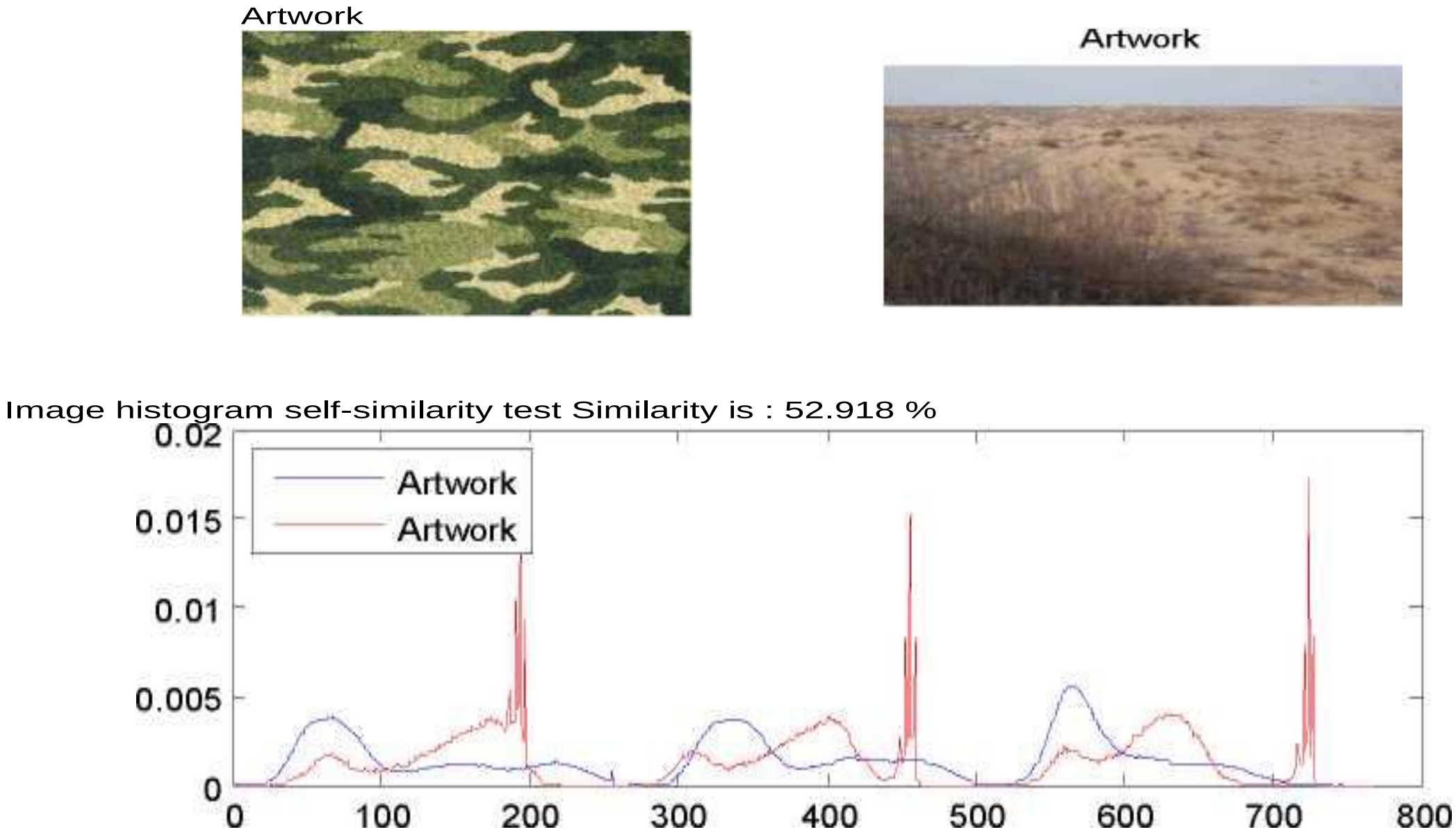

Figure2. Similarity detection of woodland camouflage with sand

From the existing similarity of woodland camouflage and sand is 52.918%, we can know that existing military camouflage performs better in the corresponding environment while the others can't adopt the environment very well. So we need synthesize four main battlefield image characteristics to design a camouflage which can adopt mostly environments.

# 4 Camouflage texture evaluation model based on gray level co-occurrence matrix

## 4.1 Similarity detection based on image texture

The image texture features comprehensive reflects the distribution and characteristics in the composition of spots, therefore, camouflage and background texture feature differences, can reflect the comprehensive characteristic differences between them. Owing to the texture is appeared repeatedly in grayscale distribution in space and form, thus a certain distance apart in the image space can exist certain gray relation between two pixels, which is the image space of gray related features. Gray level co-occurrence matrix is a kind of through the study of spatial gray level characteristics of the commonly used method to describe texture. (From baidu encyclopedia)

At co-occurrence matrix description, in order to more intuitively, derived from co-occurrence matrix four parameters reflect the status of matrix, the energy, entropy, moment of inertia, and correlation. The four characteristic parameters reflect the specific meaning is as follows:

- Energy: is the value is the sum of the squares of the gray level co-occurrence matrix elements, so also known as energy, reflects the image grey distribution uniformity degree of thickness and texture. If the co-occurrence matrix of all values are equal, the ASM is small; On the contrary, if some of the other value is big and small, ASM is bigger. When the element concentration distribution in the co-occurrence matrix, ASM value at this time. ASM value indicates that a relatively homogenous texture patterns and the rules change.
- Entropy: is the image of the information measure, also belong to the information of the image texture information, is a random measurement, when all elements in the co-occurrence matrix has the biggest randomness, all values are almost equal space co-occurrence matrix, the co-occurrence matrix elements in scattered distribution, the entropy is bigger. It says the image texture non-uniform degree or complexity.
- Relevance: it measures spatial gray level co-occurrence matrix element row or column direction on the similar degree, therefore, the relevant local gray correlation value reflects the image size. When the matrix element value equals the uniform, the value is big; On the contrary, if matrix like yuan values vary widely, it relatively small value. If the image has a horizontal grain, the horizontal direction matrix of COR COR value is greater than the rest of the matrix.
- The moment of inertia: the ability to effectively reflect the clarity of the picture.

Sheetl. Woodland's line texture feature vector value

| al | bl | a2 | b2 | a3 | b3 | a4 | b4 |
|---|---|---|---|---|---|---|---|
| 0. 0245 | 0. 0015 | 4. 0530 | 0. 0570 | 2. 3508 | 0.3114 | 0. 0794 | 0.0010 |

Sheet2. Sandy's line texture feature vector value

| al | bl | a2 | b2 | a3 | b3 | a4 | b4 |
|---|---|---|---|---|---|---|---|
| 0. 0589 | 0. 0075 | 3. 2476 | 0. 1128 | 0. 5722 | 0. 1372 | 0. 1240 | 9. 7202e —04 |

Sheet3.Desert's line texture feature vector value

| al | bl | a2 | b2 | a3 | b3 | a4 | b4 |
|---|---|---|---|---|---|---|---|
| 0.0490 | 0. 0071 | 3. 5407 | 0.1664 | 0. 9710 | 0. 3238 | 0. 0736 | 3. 7870e —04 |

Sheet4. Snow's line texture feature vector value

| al | bl | a2 | b2 | a3 | b3 | a4 | b4 |
|---|---|---|---|---|---|---|---|
| 0.0311 | 0. 0022 | 4. 1990 | 0. 0874 | 3. 7583 | 0. 7706 | 0. 0334 | 4. 6493e -04 |

# 5. The design of the camouflage based on HSV color model

### 5.1 The introduce of HSV color model:

HSV (Hue, Saturation, Value) color space is based on an intuitive nature of color created, also known as hexagonal pyramid model (Hexcone Model). The model parameters are the color: hue (H), saturation (S), Brightness (V). For people, is a more intuitive color model. Where H component represents the hue information of images, ranging from 0 ° ~ 360 °, calculated from the start in the counterclockwise direction, red is 0 °, the green is 120 °, blue for 240 °. Their complementary colors are: yellow for 60 °, blue for 180 °, magenta for 300 °; S component represents saturation information of images, ranging from

0. 0 ~ 1.0; V component represents the brightness information of images, ranging from 0.0 (black) to 1.0 (white).

In this model, on the one hand, the color component of the image is independent of V; on the other hand, H and S component and the way people feel closely related colors. So color expression and processing, HSV model than RGB model advantages. Because in its space, brightness, saturation, hue these three basic factors are independent of the operation, to reduce the complexity of the image processing, to improve the efficiency of the process.

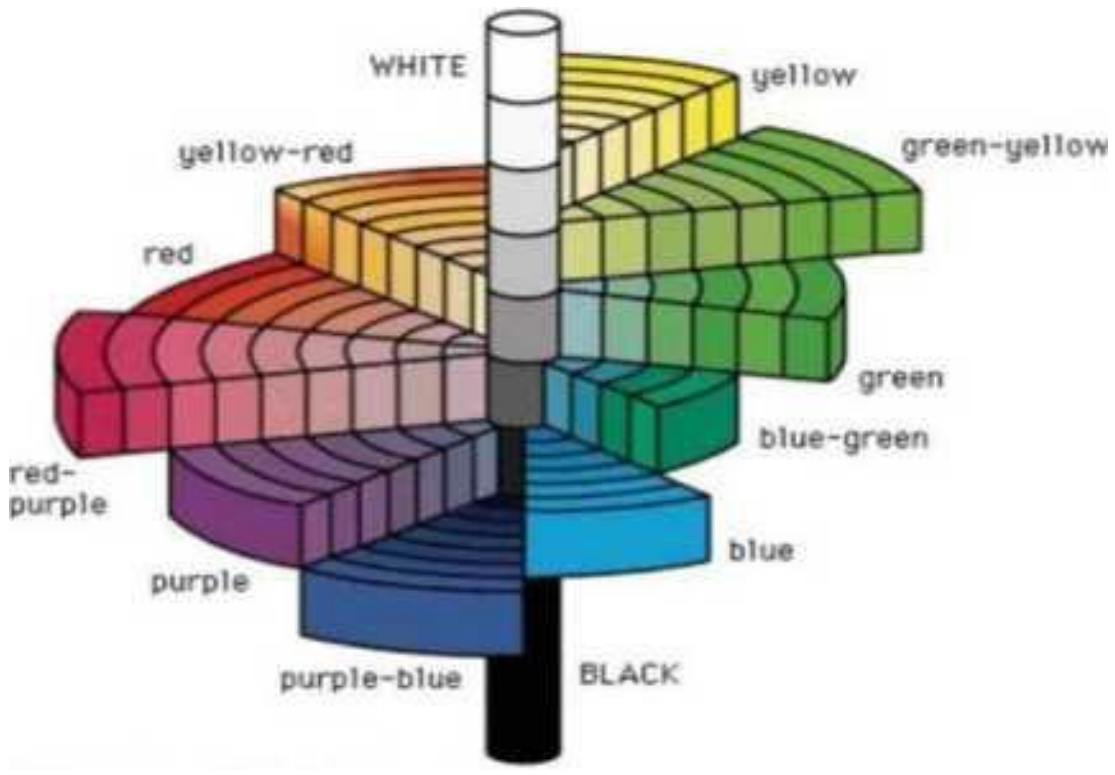

**Figure3.HSV** color model

### 5.2 The quantitative and statistical model of HSV color

Because of the similar color image pixels can be very much, or the image of the color range is very much, so the dimension of the histogram will be a lot of data volume. To simplify the model, we will appropriately HSV color space quantization. We according to human visual characteristics, the intervals will be non-quantized HSV color space, H quantized into 16, S quantization of Class 4, V quantized into four. Finally draw the histogram HS, draw the main background color woodland, sand, snow, desert four operational sites. The resulting four kinds of operational space color histogram as follows:

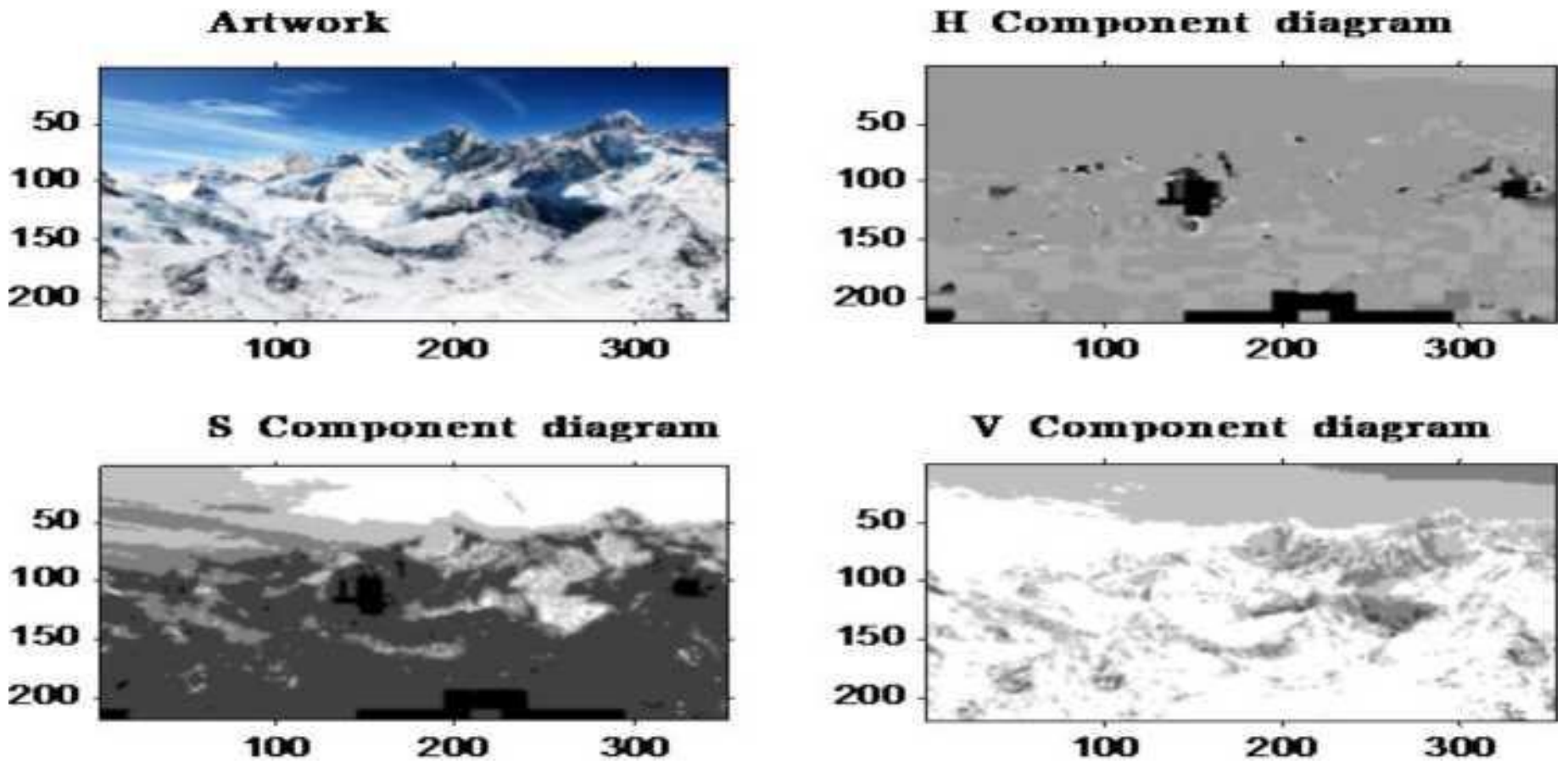

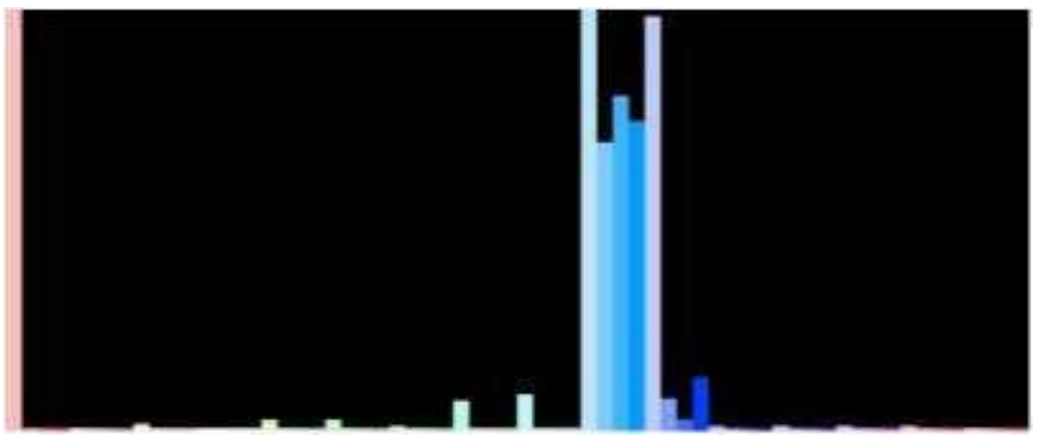

**Figure4.Snow** color histogram

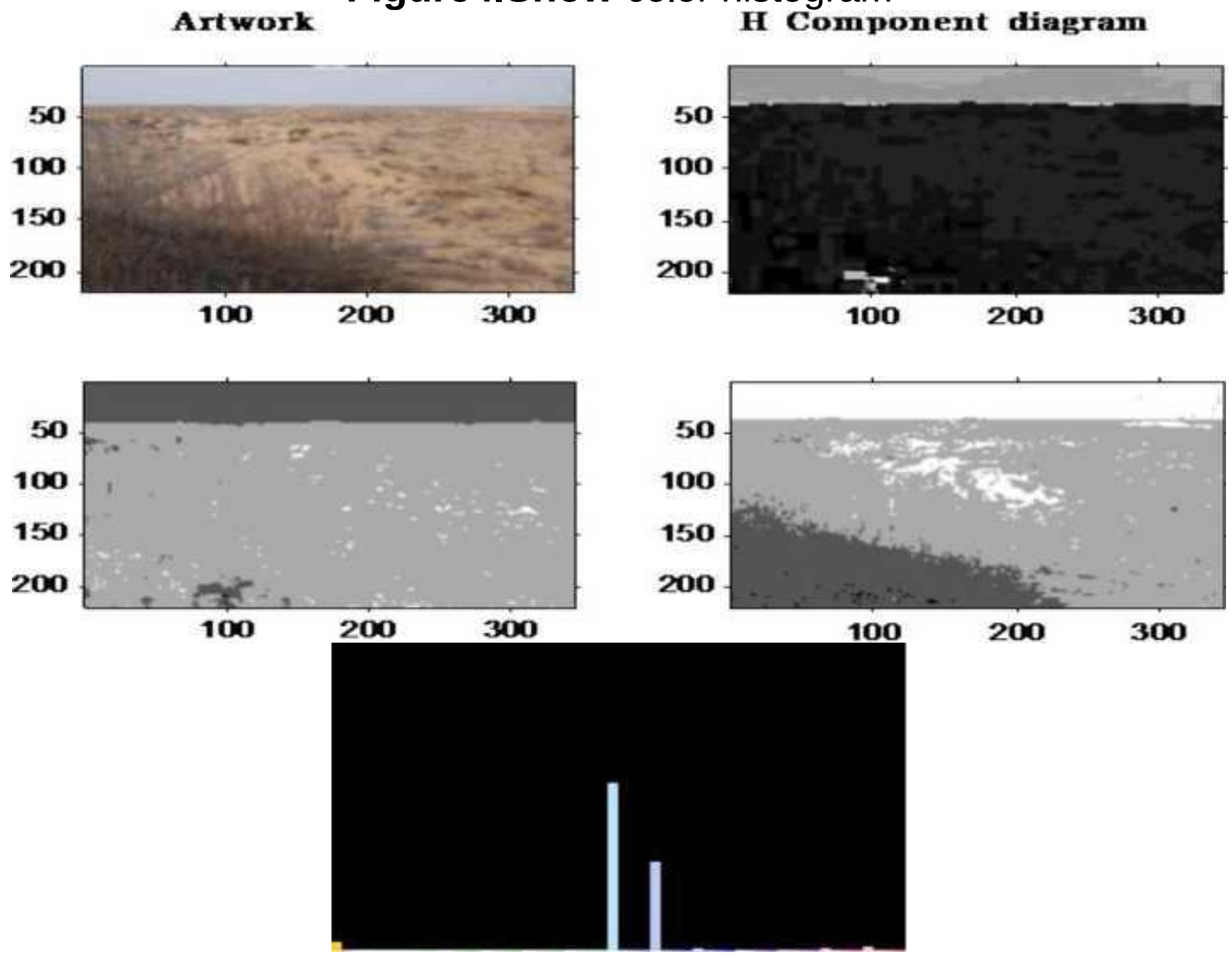

Figure5. Sandy color histogram

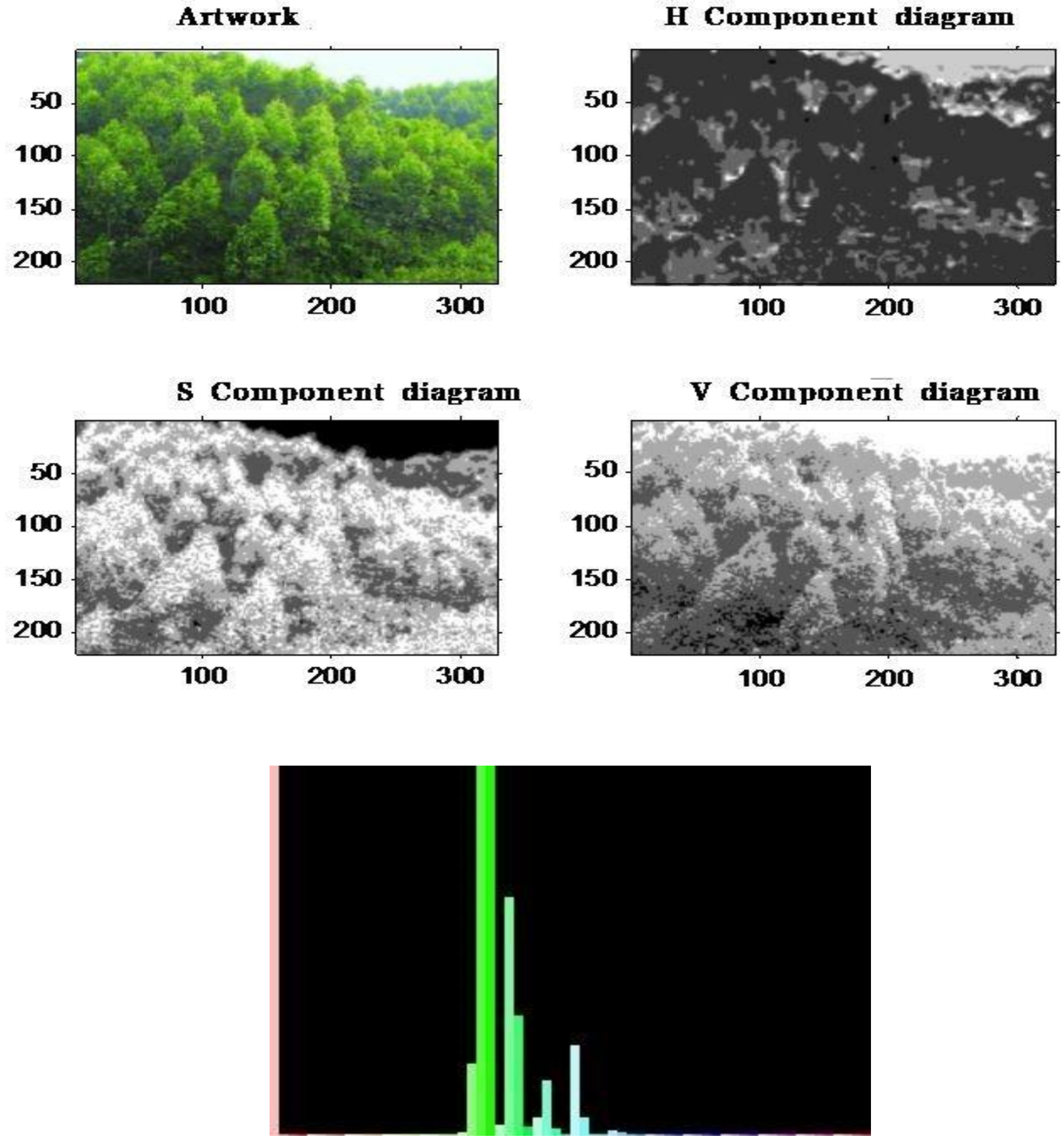

**Figure6.** Woodland color histogram

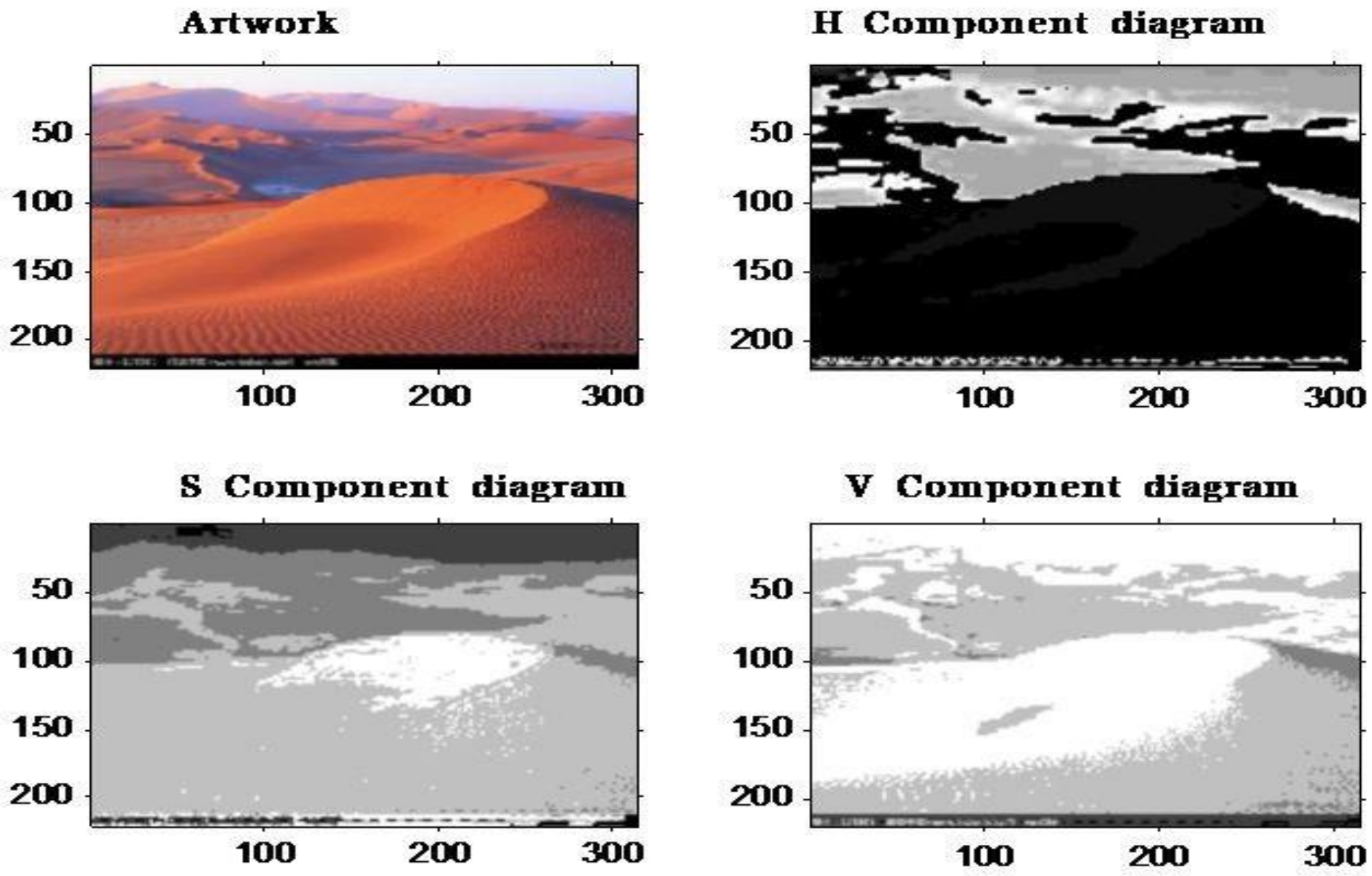

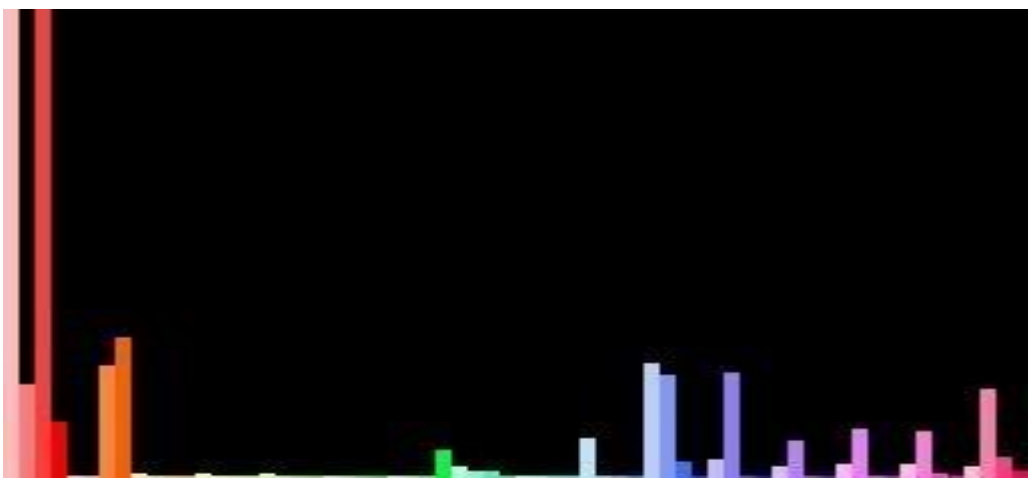

**Figure7.** Desert color histogram

# 6. Patches Design Models Based on Cluster Analysis and Edge Detection

To achieve good camouflage results, camouflage shape's choice would start from the shape of the background pattern, select the similar background pattern or similar camouflage pattern. The main contents of this section is to extract the main image patches shape method,use primarily for image segmentation techniques. In order to extract the background of the main patches shape, the method uses a K-means clustering and edge detection.

By above-mentioned method, we investigated four common goal k-means clustering analysis has been done for the background image segmentation.

As shown in the figure below:

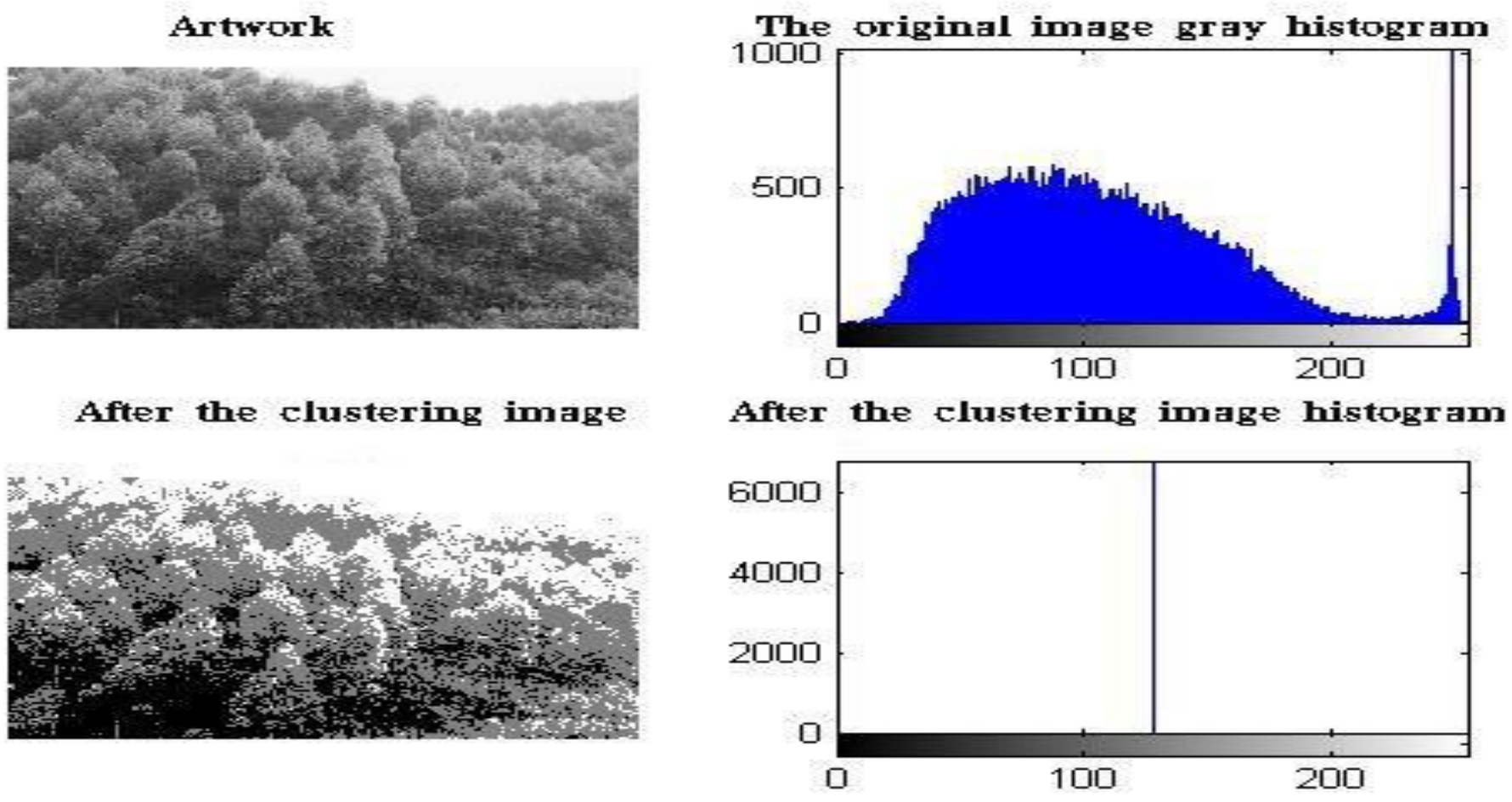

**Figure8.**Woodland clustering analysis diagram

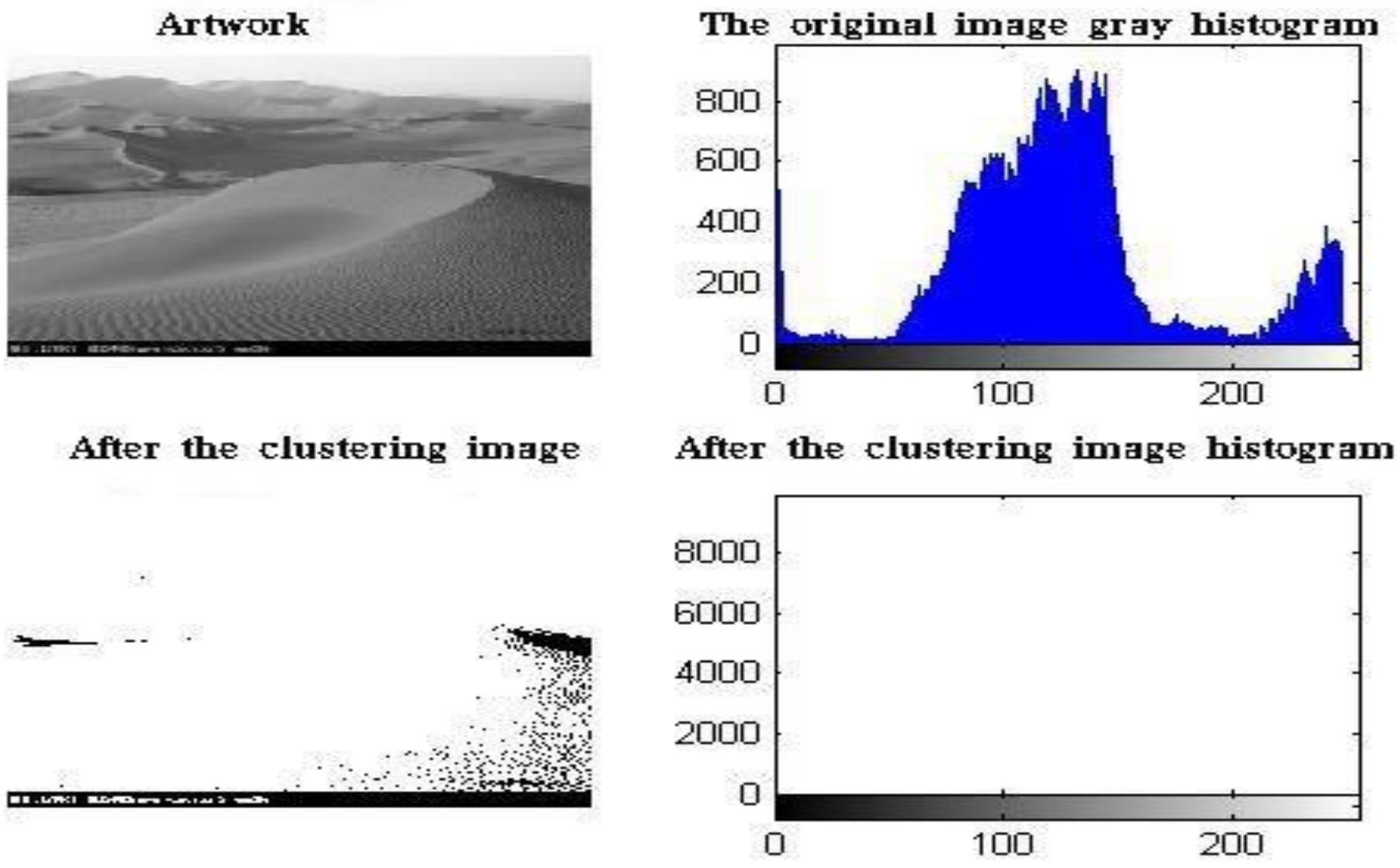

**Figure9.**Desert clustering analysis diagram

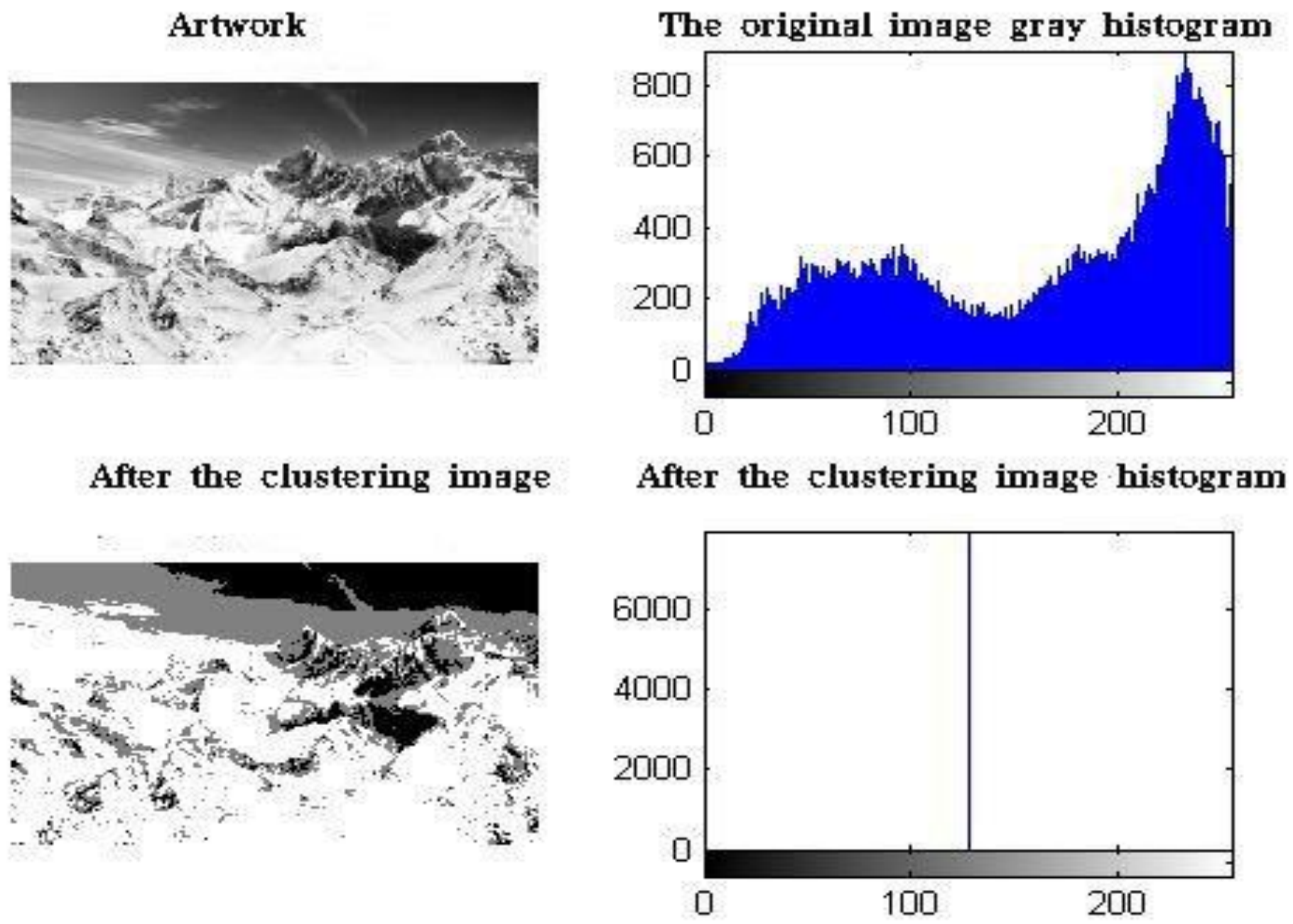

**Figure10.** Snow clustering analysis diagram

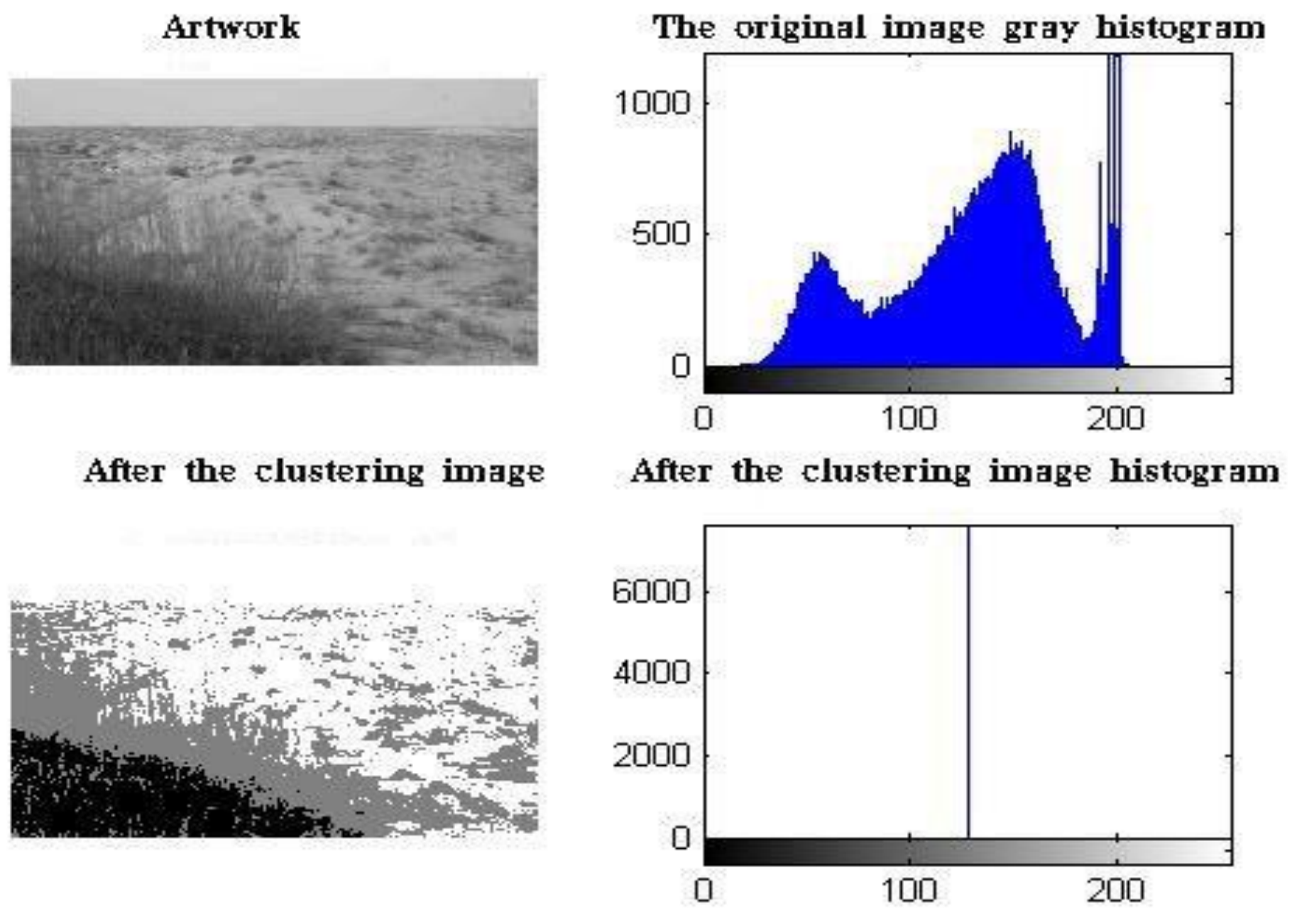

**Figure11.** Sand clustering analysis diagram

After the clustering analysis method, we made the color patches as follows:

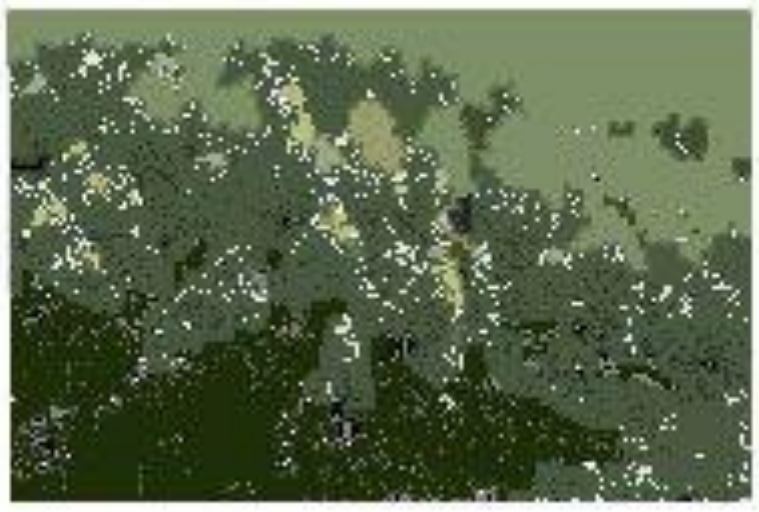 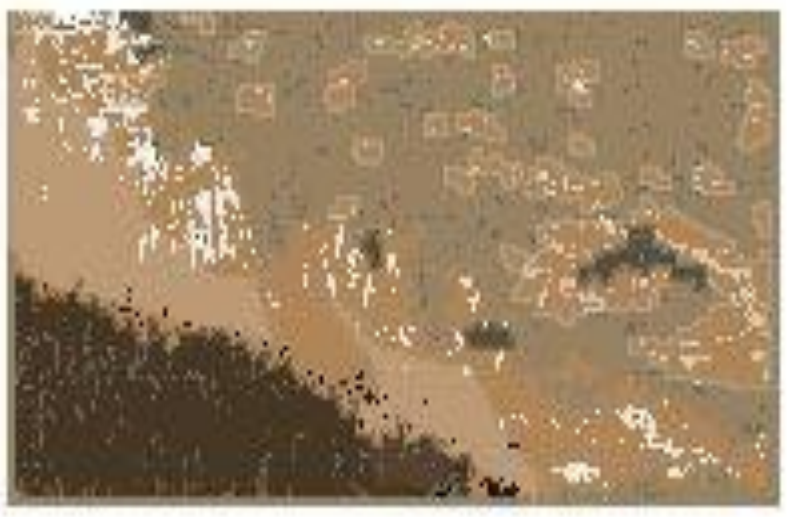

Figure12. Woodland Figure13. Sand

7.Camouflage design result and its evaluation

In order to design a camouflage for most environments, we need a comprehensive forest land, sand, desert, the four operational site the image characteristics of snow. By HSV color histogram to find the main background, we identified several main camouflage color. Through edge detection and k-means clustering analysis, we extract the contour and texture characteristics of four kinds of battlefield, thus resulting in a pattern of four kinds of camouflage. Finally comprehensive four kinds of camouflage color features and texture features, we design a kind of most universal environmental camouflage. Specific show as follows:

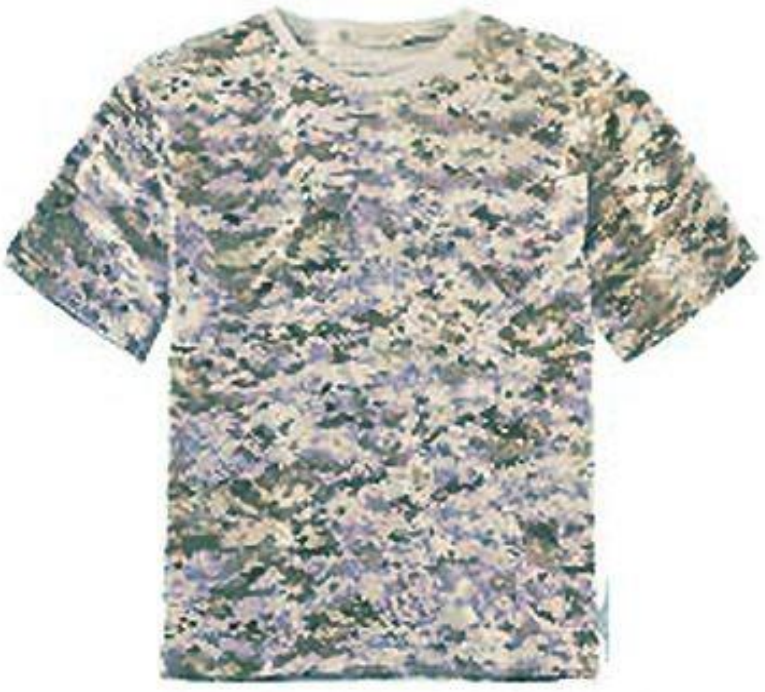

**Figure16.**Apply to most environmental camouflage pattern

Then, we will use the camouflage color and texture of the evaluation model is established to evaluate design of camouflage.

1.Camouflage and woodland picture color similarity evaluation: measured similarity was 72.5978%, that similarity is better.

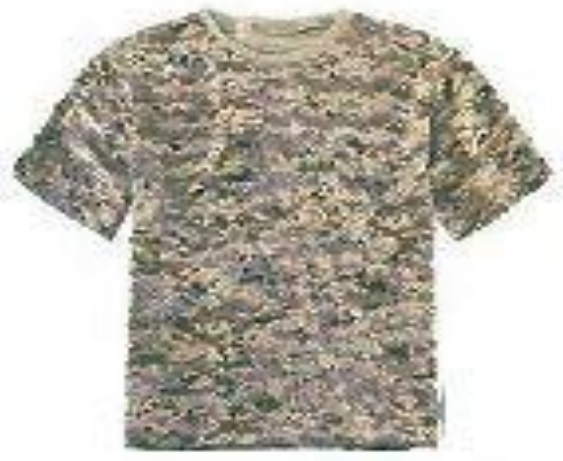 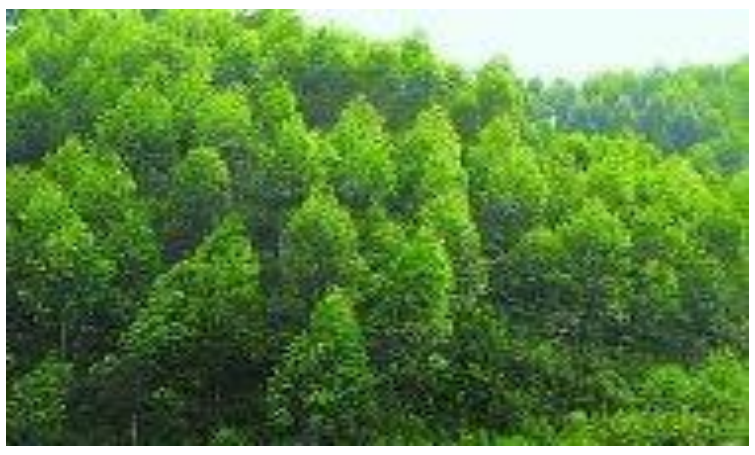

Image histogram self-similarity test Similarity is： 72.5978 %

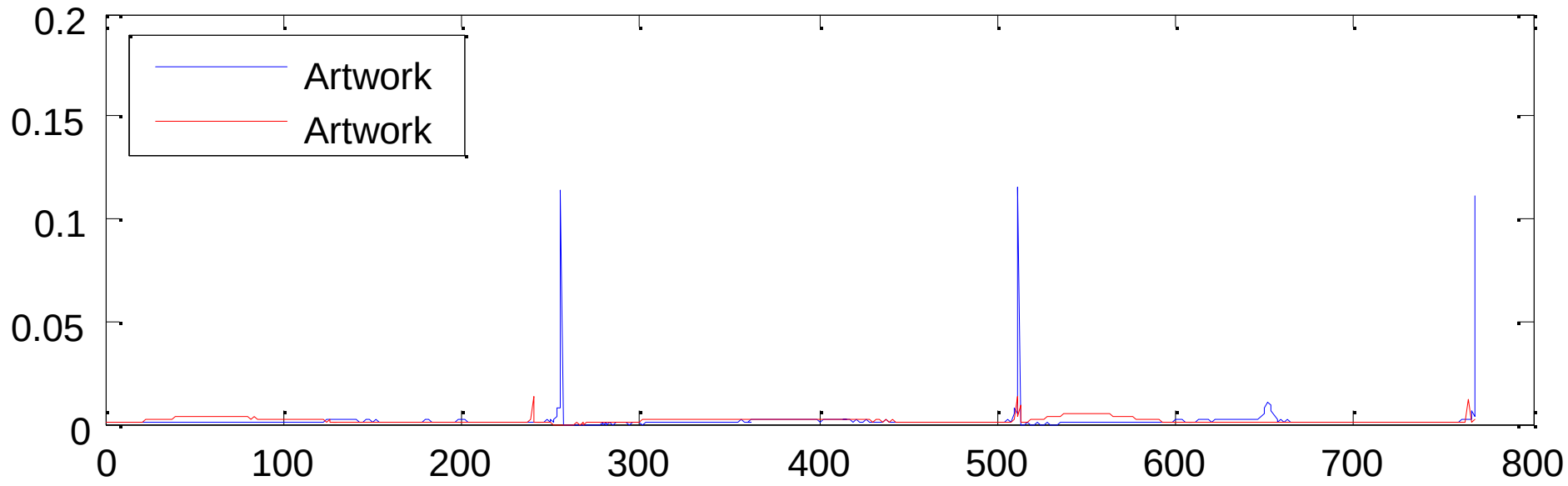

**Figure18.** Design of camouflage color similarity detection results in woodland

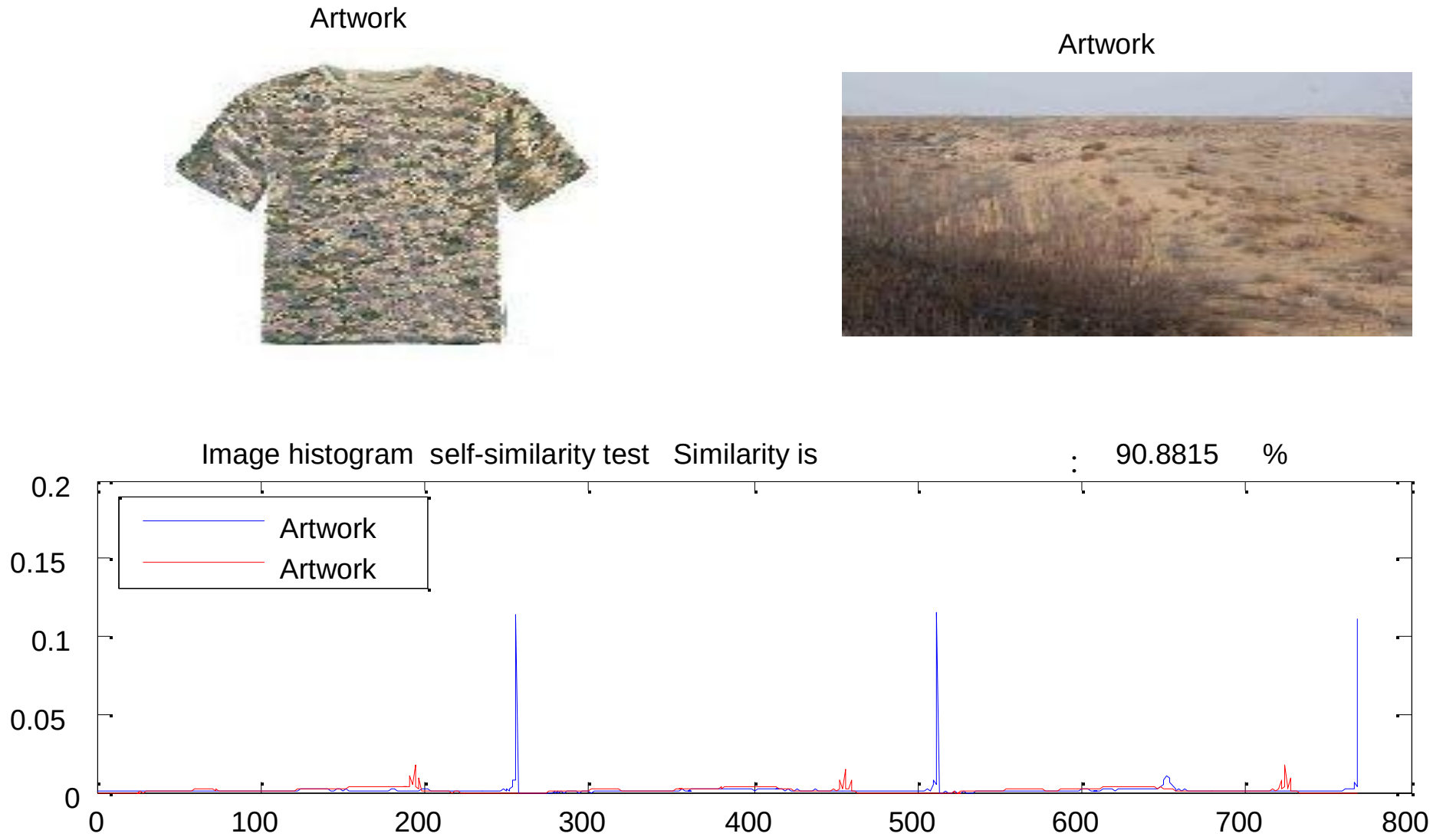

**Figure19.**Design of camouflage in the land of sand color similarity of test results

3.Camouflage and desert pictures of color similarity evaluation: measured similarity was 81.2677%, the similarity is better.

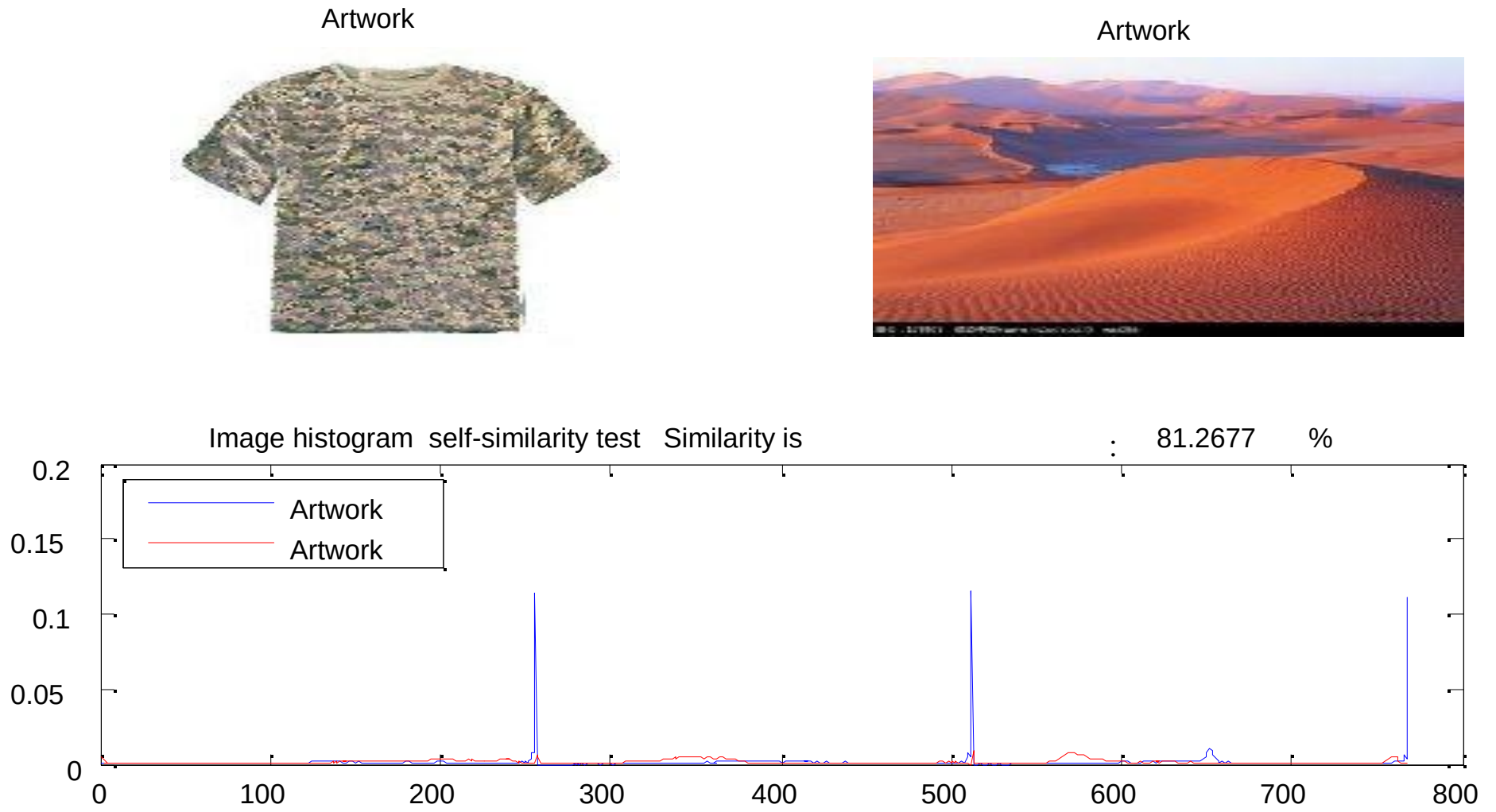

**Figure20.** In the design of camouflage in the desert color similarity test results 4.In the design of camouflage in the desert color similarity test results

Artwork

Artwork

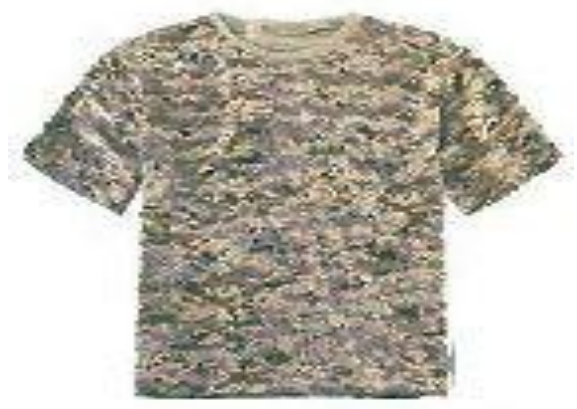

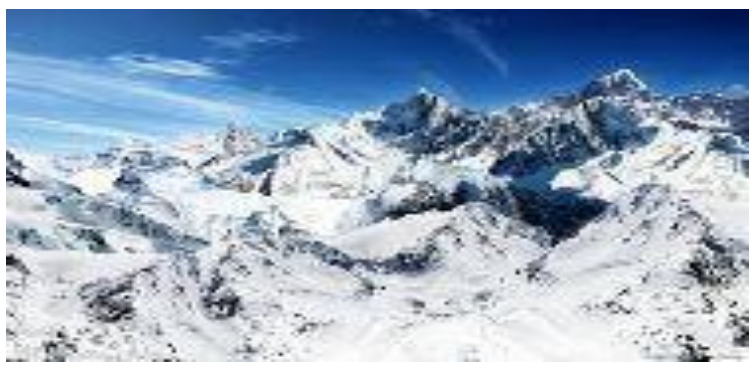

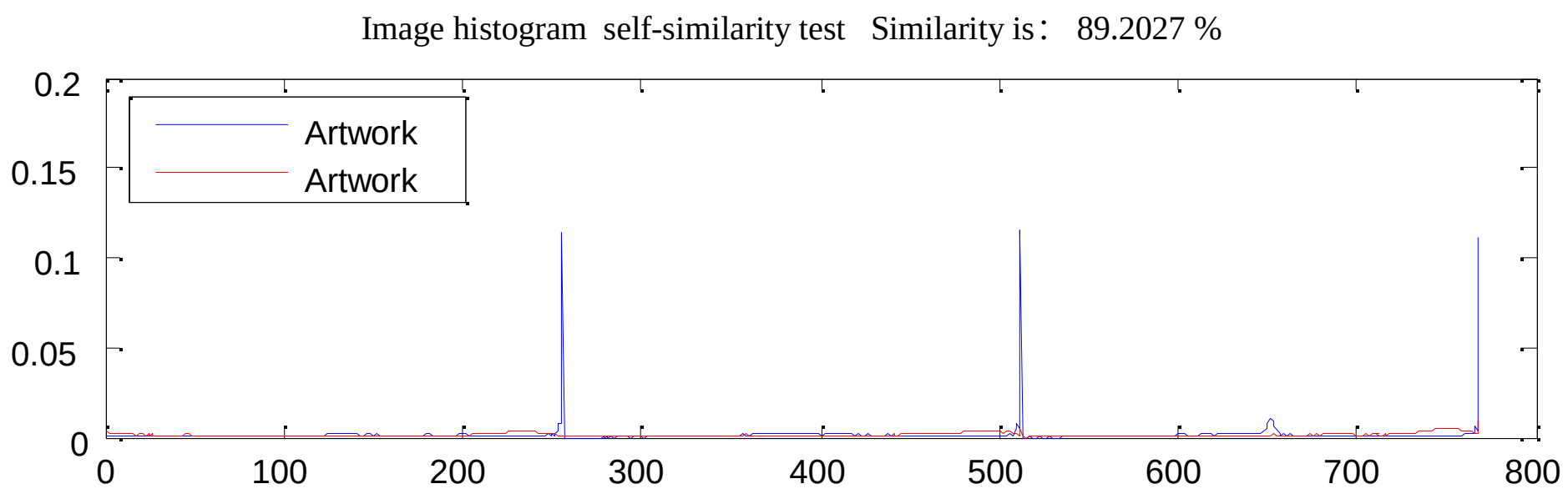

**Figure21.**Design of camouflage color similarity test results in the snow

Through the above four similarity data to prove: design of camouflage and the four kind of operational site, has good fit of the color in color, has a good camouflage effect, can apply to most of the operational environments. Sheet5.Extract design a camouflage texture parameters:

| a1 | b1 | a2 | b2 | a3 | b3 | a4 | b4 |
|---|---|---|---|---|---|---|---|
| 0.1774 | 0.0039 | 3.4041 | 0.0948 | 3.5458 | 1.0425 | 0.0591 | 0.0024 |

Through the camouflage texture characteristic parameters compared with texture feature parameters of four kinds of operational scenarios, found in camouflage texture feature parameters and the characteristics of the four scenarios were similar, all on an order of magnitude.

VIII.Website Propaganda

According to the above description, we made a static page, shows its own design and camouflage effect. Renderings are as follows:

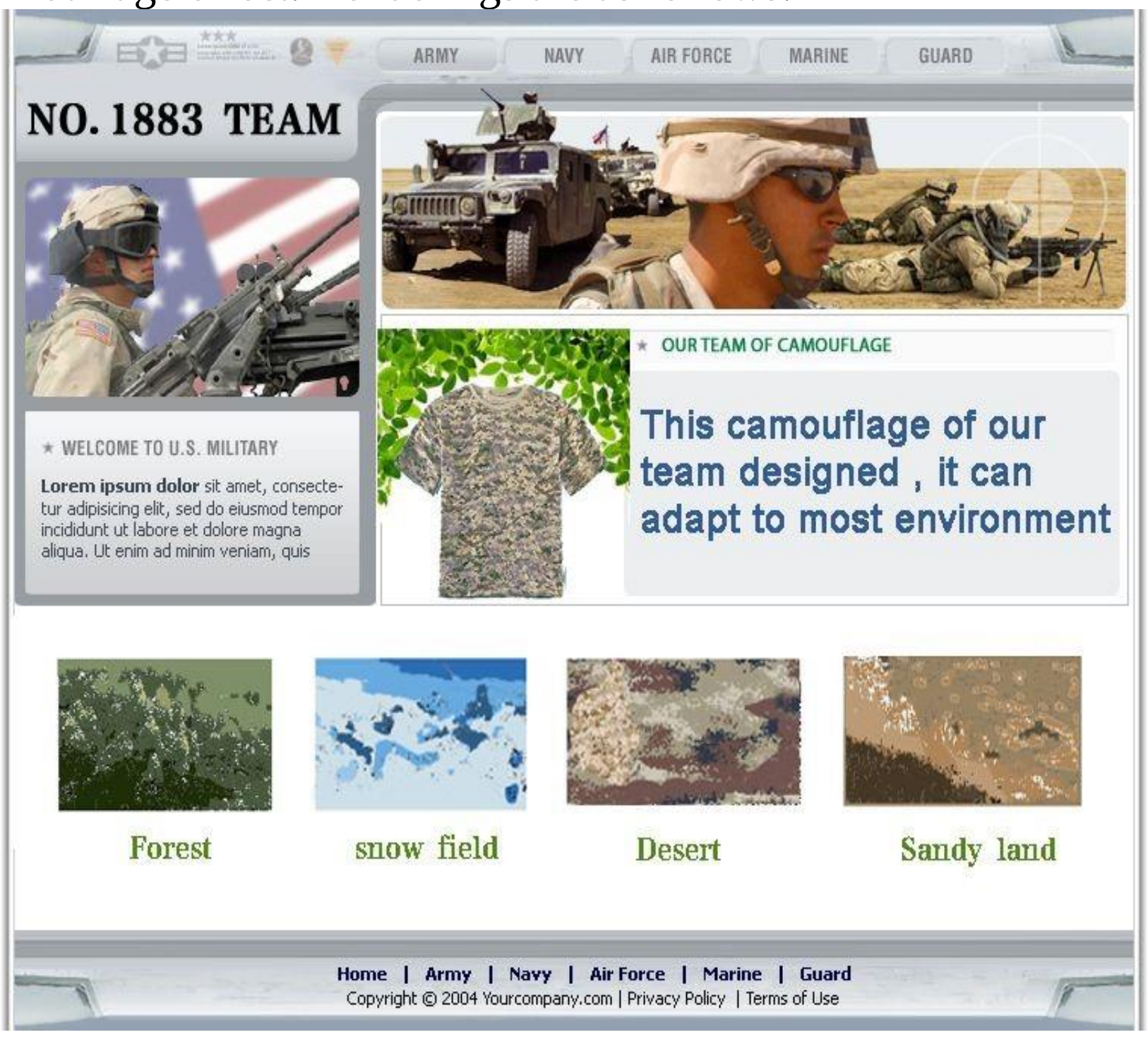

# 7. Conclusion

## Conclusions of the problem

We need to solve the problem is mainly divided into three aspects:

- Reasonable evaluation criteria, including two aspects of color similarity and texture similarity evaluation standard.
- Design can adapt to most environments of the color of camouflage. Need comprehensive image features four kind of operational site.
- Designed in most environments camouflage effect good texture pattern.

## Methods used in our models

- Based on the HSV color histogram distribution
- The texture similarity detection based on gray level co-occurrence matrix
- K-means clustering algorithm

## Applications of our models

Experiments show that our camouflage design model can adapt to most of the battle field, to overcome the existing camouflage can only adapt to the shortcoming of single combat field, increase the adaptability of camouflage. The model in terms of color and plaques reach the expected effect of camouflage, to implement the intelligent of the image feature extraction.

The model is divided into two aspects of evaluation and design, and through Matlab to extract the characteristic parameters of various background image to use Photoshop software to realize the design of camouflage, generate preview effect, convenient and quick, shorten the time of the camouflage design process.

# 8. Future Work

In this paper, the design of camouflage colors and plaque as well as evaluating the effect of camouflage, were related researches. For camouflage patch designs, including the treatment of contour extraction and edges. However, when using the K-means clustering method to extract the contour of the image, on the plaque how to match the size and mix of the effects of different ways that were not given in detail. Texture evaluation model, there is no tap more characteristic parameter, can not fully reflect the very texture of the image. In summary, we mainly work in the future is that the effect of plaque size and similarity detection image texture and contour.